\documentclass[aps,prb,reprint,groupedaddress]{revtex4-2}

\usepackage{graphicx}
\usepackage{pifont}
\usepackage{bm}
\usepackage{xcolor}
\usepackage{color}
\usepackage[colorlinks=true,linkcolor=blue,anchorcolor=blue,citecolor=blue,urlcolor=blue]{hyperref}

\begin{document}
%\begin{CJK}{UTF8}{gbsn}

%Title of paper
\title{Higher-order topological superconductivity in type-II time-reversal-symmetric Weyl semimetals with a hybrid pairing}

\author{Junkang Huang$^{1,2}$}
\author{Z. D. Wang$^{1,2,3,4}$}
\email{zwang@hku.hk}
\author{Tao Zhou$^{1,2}$}
\email{tzhou@scnu.edu.cn}

\affiliation{$^1$Guangdong Basic Research Center of Excellence for Structure and Fundamental Interactions of Matter, Guangdong Provincial Key Laboratory of Quantum Engineering and Quantum Materials, School of Physics, South China Normal University, Guangzhou 510006, China\\
	$^2$Guangdong-Hong Kong Joint Laboratory of Quantum Matter, Frontier Research Institute for Physics, South China Normal University, Guangzhou 510006, China\\
	$^3$HK Institute of Quantum Science $\&$ Technology, The University of Hong Kong, Pokfulam Road, Hong Kong, China\\
	$^4$Hong Kong Branch for Quantum Science Center of Guangdong-Hong Kong-Macau Great Bay Area, 3 Binlang Road, Shenzhen, China
}

%\date{\today}

\begin{abstract}
We employed the self-consistent method on a two-orbital type-II time-reversal-symmetric Weyl semimetal, revealing a hybrid pairing of singlet $s$-wave and triplet $p$-wave. We present a detailed analysis of the normal-state electronic structure and the self-consistent results. Our findings indicate that the selection of hybrid pairings is governed by distinct surface Fermi-arc configurations: specifically, $s$-wave pairing dominates on the bottom surface, while $p$-wave pairing prevails on the top. Furthermore, the emergent superconducting state is a second-order topological superconductors with hinge states in the system. Our results identify type-II time-reversal-invariant Weyl semimetals as a promising intrinsic platform for realizing unconventional and topological superconductivity.
\end{abstract}
\maketitle

\section{\label{Intro}Introduction}
Superconductivity in topologically non-trivial materials represents one of the most compelling frontiers in condensed matter physics, where the interplay between exotic electronic structures and quantum phenomena gives rise to exotic physical properties. Among these systems, Weyl semimetals (WSM) have emerged as particularly intriguing candidates, characterized by their paired Weyl nodes and topological Fermi arc. The fundamental question driving current research is whether these unique electronic configurations and nontrivial topological properties can induce unconventional or topological superconductivity—a question that has motivated numerous experimental investigations in the past \cite{veyrat_berezinskii_2023,cao_pressureinduced_2022,wang_visualizing_2021,wang_pressureinduced_2021,emmanouilidou_fermiology_2020,tang_strongcoupling_2021,shang_spintriplet_2022}. 

Time-reversal-symmetric (TR) WSM represent a more promising platform for superconductivity compared with their TR-breaking counterparts, since superconductivity is suppressed by the breaking of TR. This advantage has led to the discovery of superconductivity in numerous TRWSM have been reported to exhibit the signature of superconductivity, such as type-I TRWSM like PtBi$_2$~\cite{kuibarov_three_2025,shipunov_polymorphic_2020,zabala_enhanced_2024}, NbX (X = As,P)~\cite{bachmann_inducing_2017,deng_pressurequenched_2022}, TaX (X = As,P)~\cite{wang_discovery_2017,hou_two_2020,luo_surface_2019}, and type-II TRWSM including MoTe$_2$~\cite{qi_superconductivity_2016,chen_superconductivity_2016}, WTe$_2$~\cite{pan_pressuredriven_2015,kang_superconductivity_2015} and TalrTe$_4$~\cite{xing_surface_2020,cai_observation_2019}.

The exotic electronic structure of TRWSM, particularly the surface-localized Fermi arcs, has been shown to play a crucial role in their superconducting behavior. Previously, edge supercurrents have been detected in MoTe$_2$ \cite{wang_evidence_2020}. When proximized to a conventional $s$-wave superconductor, the observed supercurrent indicates an incompatibility between the intrinsic superconductivity and the conventional superconductivity. These two findings imply the possible existence of intrinsic unconventional topological superconductivity in MoTe$_2$ \cite{kim_edge_2024}. Additionally, muon-spin relaxation and rotation ($\mu$SR) experiments have revealed an unconventional superconducting pairing order in $T_d$-MoTe$_2$ \cite{guguchia_signatures_2017}.

More compelling evidence for unconventional superconductivity emerges in PtBi$_2$ compounds. Recent angle-resolved photoemission spectroscopy (ARPES) experiments on PtBi$_2$ compounds have provided direct evidence for superconducting Fermi arcs and revealed an anisotropic superconducting gap structure, strongly suggesting unconventional pairing mechanisms \cite{kuibarov_evidence_2024,changdar_topological_2025}. Complementary STM measurements have quantified the superconducting gaps of approximately 20 meV in t-PtBi$_2$ \cite{schimmel_surface_2024a,huang_sizable_2025} and 0.5 meV in $\gamma$-PtBi$_2$~\cite{moreno_robust_2025}, respectively. Theoretical studies on superconductivity in TRWSM mostly concentrate on surface superconductivity in type-I systems \cite{waje_ginzburglandau_2025,trama_selfconsistent_2025,vocaturo_electronic_2024,bai_superconductivity_2025,nomani_intrinsic_2023}, topological properties \cite{giwa_superconductor_2023,giwa_fermi_2021,zhang_higherorder_2020,yan_vortex_2020,hamilton_topological_2019,hosur_timereversalinvariant_2014}, and Andreev reflection phenomena at normal-superconductor interfaces~\cite{azizi_andreev_2020}.

In the past several years, the emergence of second-order topological three-dimensional superconductors have attracted considerable interest, characterized by hinge states. Such systems have been theoretically proposed to be realized through chiral superconducting pairing \cite{khalaf_higherorder_2018,fu_chiral_2021} or via heterostructures composed of topological insulators and conventional superconductors \cite{hsu_majorana_2018,wu_inplane_2020}. Recent theoretical work has suggested that PtBi$_2$ may exhibit chiral superconductivity accompanied by hinge states \cite{changdar_topological_2025}.
Type-II TRWSM possess tilted Weyl cones along with numerous Fermi arcs facilitating the formation of large Fermi surface \cite{mccormick_minimal_2017}, which is conducive to superconductivity that may further give rise to unconventional topological superconductivity. However, theoretical investigations into this promising platform remain relatively limited.

In our paper, a two-orbital tight-binding model is presented to describe the type-II TRWSM. The electronic structure has been meticulously studied. We employ the self-consistent method to obtain the superconducting-state ground state. The resulting pairing in our model is a hybrid pairing of singlet $s$-wave and triplet $p$-wave pairings. 
The system's large zero-energy density of states and abundant surface Fermi arcs enable this superconducting phase to emerge with relatively weak pairing potentials compared to time-reversal-breaking or type-I TR system \cite{zhou_superconductivity_2016,bai_superconductivity_2025}.
The resulting bulk superconductor is fully gapped and possesses non-trivial second-order topological hinge states. This higher-order character is confirmed by the local electronic structure and quadrupole moment. Our work intends to provide new insights into unconventional and topological superconductivity in the type-II TRWSM.

The remainder of this paper is organized as follows: Section~II presents the model and the formalism; Section~III discusses the numerical results; and Section~IV provides a summary.

\section{\label{sec:Model}Model and Formalism}

We consider a minimal Hamiltonian for TRWSM expressed as follows:
\begin{eqnarray}
&& H = \sum_{{\bf k}\tau\sigma} \left( \gamma\cos2{k_x} \cos{k_y} \right)c_{{\bf k} \tau\sigma}^{\dagger}c_{{\bf k}\tau\sigma} + \sum_{{\bf k}\sigma} \left[ \left( M - 2t_x\cos{k_x} \right. \right. \nonumber \\
&& \left. - 2t_y\cos2{k_y} - 2t_z\cos{\bf k_z} \right)\left( c_{{\bf k} \tau\sigma}^{\dagger}c_{{\bf k}\tau'\sigma} + c_{{\bf k} \tau'\sigma}^{\dagger}c_{{\bf k}\tau\sigma} \right) \nonumber \\
&& + 2\lambda_y\cos{k_y}\left( c_{{\bf k} \tau\sigma}^{\dagger}c_{{\bf k}\tau\sigma} - c_{{\bf k} \tau'\sigma}^{\dagger}c_{{\bf k}\tau'\sigma} \right) \nonumber \\
&& \left. + 2\lambda_z\sin{\bf k_z}\left( -ic_{{\bf k} \tau\sigma}^{\dagger}c_{{\bf k}\tau'\sigma} + ic_{{\bf k} \tau'\sigma}^{\dagger}c_{{\bf k}\tau\sigma} \right) \right] + H_{I},
\end{eqnarray}
where \( \tau/\tau' \) and \( \sigma \) denote the orbital and spin indices, respectively. $M$ denotes the onsite interorbital coupling constant. $t_x$, $t_y$, $t_z$ and $\lambda_z$ represent the offsite interorbital coupling constants. $\lambda_y$ signifies the intraorbital hopping constant. Notably, both $\lambda_y$ and $\lambda_z$ are responsible for inducing the non-trivial topological properties of system. The transition from type-I to typle-II WSM is governed by the long-range intraorbital constant $\gamma$.
This model hosts two pairs of Weyl points located at $\left( \pm\pi/2, \pm\pi/2, 0 \right)$. For $\left|\gamma\right| < 0.5$, the system behaves as a type-I Weyl semimetal, characterized by a Fermi surface comprising only the four Weyl nodes. When $\left|\gamma\right| > 0.5$, it transitions into a type-II Weyl semimetal. At this time, where the emergence of electron and hole Fermi pockets leads to an extended Fermi surface.
$H_I$ is the nearest neighbor intraorbital attractive exchange interaction term, which accounts for the superconducting pairing, being given by:
\begin{eqnarray}
H_I = -V\sum_{i\alpha} n_{i,\tau\uparrow}n_{i+\alpha,\tau\downarrow}.
\end{eqnarray}
Here, $n_{i,\tau\uparrow}$ represents the particle number operator, and $n_{i,\tau\uparrow} = c_{i,\tau\uparrow}^{\dagger}c_{i,\tau\uparrow}$. The mean-field order parameters can be defined as $\Delta_{\tau\alpha} = V \langle c_{i,\tau\uparrow} c_{i+\alpha,\tau\downarrow} \rangle$, where $\alpha$ denotes the six nearest neighbor vectors. $V$ represents the superconducting pairing strength.

For our model, the Fermi arcs are localized in the surfaces of the system and can be exuded by opening the boundary in the $z$ direction. In this partial opening boundary condition, the Hamiltonian can be expressed as $H = \sum_{z,{\bf k}}\Psi^{\dagger}_{z,{\bf k}}\hat{M}_z\left( \bf k \right)\Psi_{z,{\bf k}}$, where $\hat{M}_z$ is a $2N_z\times2N_z$ matrix and the Nambu vector $\Psi_{z,{\bf k}} = \left( u_{z,{\bf k}\tau},  v_{z,{\bf k}\tau} \right)^T$. $u_{z,{\bf k}\tau} = \left( c_{z,{\bf k}1\uparrow}.c_{z,{\bf k}2\uparrow} \right)$ and $v_{z,{\bf k}\tau} = \left( c_{z,{-\bf k}1\downarrow}^{\dagger}, c_{z,{-\bf k}2\downarrow}^{\dagger} \right)$. Here the indices 1 and 2 denote the orbitals. $z/z'$ is the layer index.

The superconducting pairing orders are not independent and should be determined by the other parameters in the model. In our paper, the order parameters are obtained by self-consistent calculation. The self-consistent equations are expressed as:
\begin{eqnarray}
&&\Delta_{\tau\alpha'}^z = \frac{V}{2N_1}\sum_{{\bf k}\tau n}u_{z,{\bf k}\tau n}v_{z,{\bf k}\tau n}^*\tanh\left( \frac{\beta E_{{\bf k}n}}{2} \right)\exp\left( -i{\bf k}\cdot\alpha' \right), \nonumber \\
\\
&&\Delta_{\tau}^{z,z'} = \frac{V}{2N_1}\sum_{{\bf k}\tau n}u_{z,{\bf k}\tau n}v_{z',{\bf k}\tau n}^*\tanh\left( \frac{\beta E_{{\bf k}n}}{2} \right),
\end{eqnarray}
where $\alpha'$ denotes the four in-plane nearest-neighbor vectors $\hat{x}$, $\hat{y}$, $-\hat{x}$ and $-\hat{y}$. $\Delta_{\tau\alpha'}^z$ is the intralayer pairing order parameter in the $z$ layer. $\Delta_{\tau}^{z,z'}$ is the interlayer pairing order parameter from $z$ to $z'$ layers ($\left|z - z'\right| = 1$). $N_1$ is the number of unit cells in ${k_x}$-${k_y}$ momentum space.

The spectral function of layer $z$ and orbital $\tau$ can be calculated by:
\begin{eqnarray}
A_{\tau}^z\left( {k_x},{k_y},\omega \right) = -\frac{1}{\pi}\sum_n\mathrm{Im}\frac{\left| u_{2z+\tau-2,n}\left( {k_x},{k_y} \right) \right|^2}{\omega - E_n\left( {k_x},{k_y} \right) + i\Gamma},
\end{eqnarray}
The orbital index $\tau = 1$ and $\tau = 2$ denote the orbital $1$ and $2$, respectively. The eigenvector $u_{2z+\tau-2,n}\left( {k_x},{k_y} \right)$ and eigenvalue $E_n\left( {k_x},{k_y} \right)$ are obtained by diagonalizing the Hamiltonian with the partially opening boundary condition in the $z$ direction. $\Gamma$ is a damping factor being set as $\Gamma = 0.01$.

The local density of states (LDOS) of layer $z$ and orbital $\tau$ can be expressed by integrating the spectral function $A_{\tau}^z\left( {k_x},{k_y},\omega \right)$ over the entire ${k_x}$-${k_y}$ momentum space:
\begin{eqnarray}
\rho_{\tau}^z\left(\omega\right) = \frac{1}{N_1}\sum_{{k_x},{k_y}} A_{\tau}^z\left( {k_x},{k_y},\omega \right).
\end{eqnarray}

To explore the hinge states in the system, we have considered the Hamiltonian with the partially opening boundary condition in the $x$ and $z$ direction. In this boundary condition, the Hamiltonian is given by $H = \sum_{xz,{k_y}}\Psi^{\dagger}_{xz,{k_y}}\hat{M}_{xz}\left( \bf k_y \right)\Psi_{xz,{k_y}}$. $\hat{M}_{xz}$ is a $2N_xN_z\times2N_xN_z$ matrix.

The $\tau$-orbital spectral function at site $\left(x,z\right)$ is expressed as:
\begin{eqnarray}
A_{\tau}^{\left(x,z\right)}\left( {k_y},\omega \right) = -\frac{1}{\pi}\sum_n\mathrm{Im}\frac{\left| u_{2N_xz+2x+\tau-2N_x-2,n}\left( {k_y} \right) \right|^2}{\omega - E_n\left( {k_y} \right) + i\Gamma}, \nonumber \\
\end{eqnarray}

For the three-dimensional system, we can define the quadrupole moment $q_{xz}\left({k_y}\right)$ in $x$-$z$ plane for each ${k_y}$-sliced layer~\cite{fu_chiral_2021}. The quadrupole moment $q_{xz}\left({k_y}\right)$ can be expressed as:
\begin{eqnarray}
q_{xz}\left({k_y}\right) = \frac{1}{2\pi}\mathrm{Im}\log\left[ \mathrm{Det}\left( \Phi_{k_y}^{\dagger} Q \Phi_{k_y}\right) \sqrt{\mathrm{Det}\left(Q^{\dagger}\right)} \right].
\end{eqnarray}
Here $\Phi_{k_y}$ is a $2N_xN_z \times N_xN_z$ matrix, which is constructed by columnwise packing the occupied eigenstates with $\Phi_{k_y} = \left( u_{1,{k_y}},u_{2,{k_y}},...,u_{N_xN_z,{k_y}} \right)$. The matrix $Q$ is a $2N_xN_z \times 2N_xN_z$ diagonal constant matrix. The diagonal elements are defined as $e^{2\pi ixz/N_xN_z}$.

In the following presented results, unless otherwise specified, the input parameters are set as $M = 1$, $t_x = t_z = \lambda_y = \lambda_z = 1$, $t_y = 0.5$, $\gamma = 1$, $N_x = N_z = 50$ and the pairing strength $V = 2$.

\section{Results and Discussion}
\subsection{Normal-state electronic structure}
\begin{figure}
\centering
\includegraphics[width = 9cm]{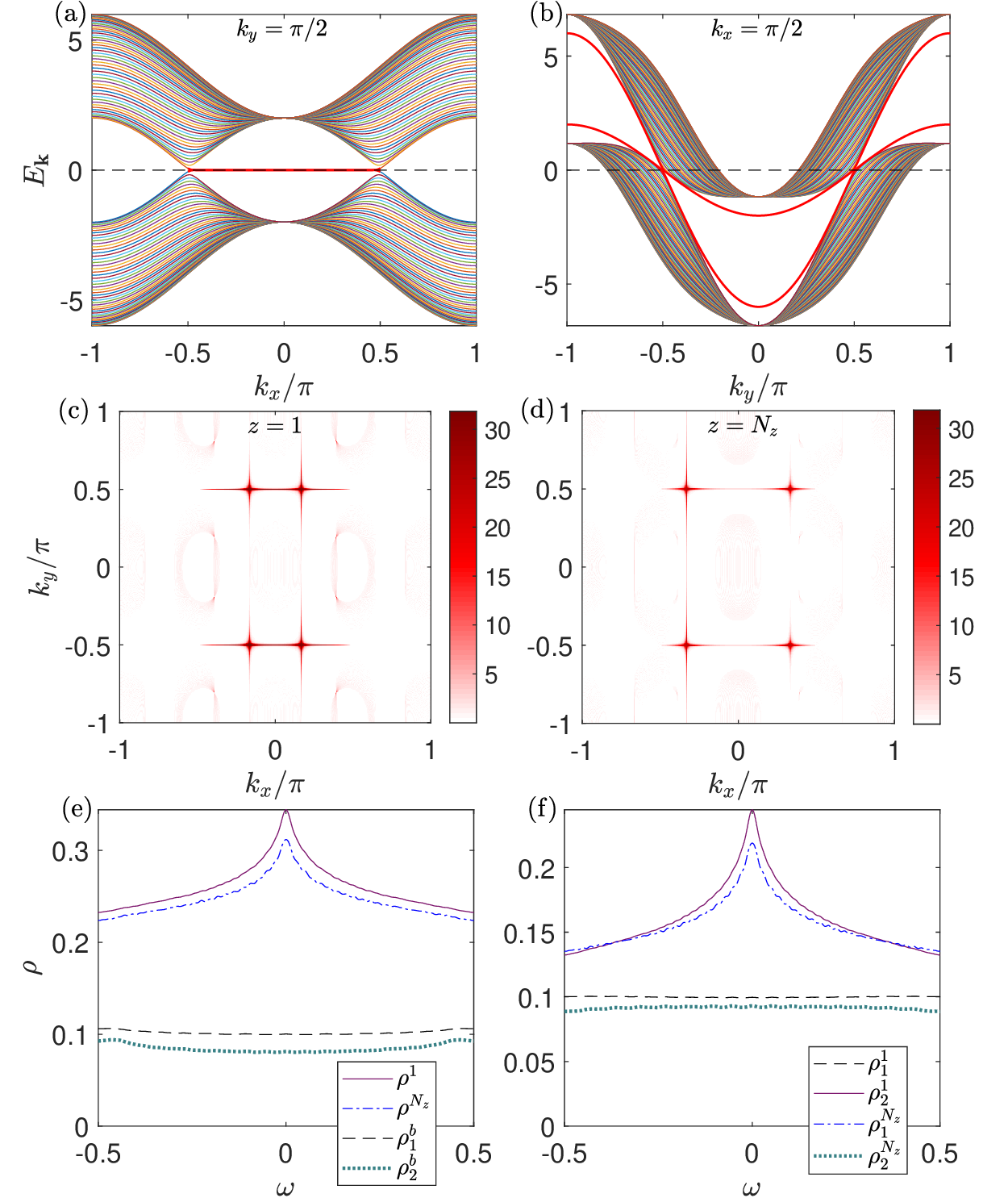}
\caption{\label{Fig1} Normal-state electronic structures with partially opening boundary condition in the $z$ direction. Top row: energy spectra at the ${k_y} = \pi/2$ slice (a), and at the ${k_x} = \pi/2$ slice (b). Middle row: zero-energy spectral functions on the $z = 1$ surface (c) and on the $z = N_z$ surface (d). Bottom row: orbital-resolved LDOS curves on the $z = 1$, $z = N_z$ surfaces and in the bulk ($z = N_z/2$).}
\end{figure}

We begin by presenting the normal-state electronic structure. Fig.~\ref{Fig1} summarises the bulk and surface energy spectra with the partially opening boundary condition in the $z$ direction. The energy spectrum for ${k_y} = 0.5\pi$ slice is shown in Fig.~\ref{Fig1}(a). The double-degenerate Fermi arcs with bold red lines match two Weyl points $\left( \pm\pi/2,\pi/2 \right)$ carrying opposite chirality. At the ${k_x}$ direction, the Weyl cones are not tilted by $\gamma$ term.
The energy spectrum for ${k_x} = 0.5\pi$ slice is represented in Fig.~\ref{Fig1}(b). The oblique edge bands with bold red lines intersect at the locations of two Weyl points. Each tilted Weyl cone generates an electron pocket and a hole pocket at the two sides of the Weyl points. The electron and hole pockets are connected by the Weyl points.
The tilted Weyl cones form large Fermi surface inducing numerous zero-energy electronic states, which contrasts to conventional Type-I WSM. Time-reversal-symmetry, large Fermi surface and numerous zero-energy electronic states are beneficial to the superconductivity.

The zero-energy spectral functions on the two surfaces of the WSM ($A^{1}$ and $A^{N_z}$) are exhibited in Figs.~\ref{Fig1}(c) and \ref{Fig1}(d), where $A^{z}$ represents the spectral function for the both two orbital channels, as $A^{z} = A_1^{z} + A_2^{z}$. On both surfaces, the horizontal and vertical Fermi arcs contribute most of zero-energy states, while the zero-energy states projected by the bulk are much weaker. Due to the broken inversion symmetry, the locations of Fermi arcs on the two surfaces are different.
On the $z = 1$ surface, as shown in Fig.~\ref{Fig1}(c), each shorter ${k_y} = \pm 0.5\pi$ horizontal Fermi arc connect two Weyl points. The longer vertical Fermi arcs intersect with the shorter ones at points $\left(\pm0.16\pi,\pm0.5\pi\right)$. The longer Fermi arcs do not connect Weyl points and stretch across the Brillouin zone in the ${k_y}$ direction. Furthermore, the two longer Fermi arcs are very close together.
On the $z = N_z$ surface, presented in Fig.~\ref{Fig1}(d), the ${k_y} = \pm 0.5\pi$ horizontal Fermi arcs are shorter and connect two Weyl points, while ${k_x} = \pm0.33\pi$ vertical Fermi arcs are longer and stretch across the Brillouin zone. Compared with the longer vertical Fermi arcs on the $z = 1$ surface, the vertical ones on the $z = N_z$ surface are separated far away.

The contribution of Fermi arcs can be unveiled by the LDOS curves, showcased in Figs.~\ref{Fig1}(e) and \ref{Fig1}(f). $\rho^1$ and $\rho^{N_z}$ without subscript signify $\rho^{z} = \rho^{z}_1 + \rho^{z}_2$.
In Fig.~\ref{Fig1}(e), the Fermi arcs induce a sharp zero-energy peak localized on the surface $z = 1$ (solid crimson line) and surface $z = N_z$ (blue dash-dotted line). The bulk LDOS for orbitals 1 and 2 is weaker and flatter at low-energy level. From Figs.~\ref{Fig1}(e) and \ref{Fig1}(f), it is found that
the orbital 2 dominates the localized sharp zero-energy peaks on the $z = 1$ surface, while orbital 1 dominates the sharp peaks on the $z = N_z$ surface, due to the inversion symmetry being broken.
In Fig.~\ref{Fig1}(f), LDOS for orbital 1 on the surface $z = 1$ is plotted as a black dashed line, and that for orbital 2 on the surface $z = N_z$ with a green dotted line are weak and almost projected by the bulk LDOS.

\begin{figure}
\centering
\includegraphics[width = 9cm]{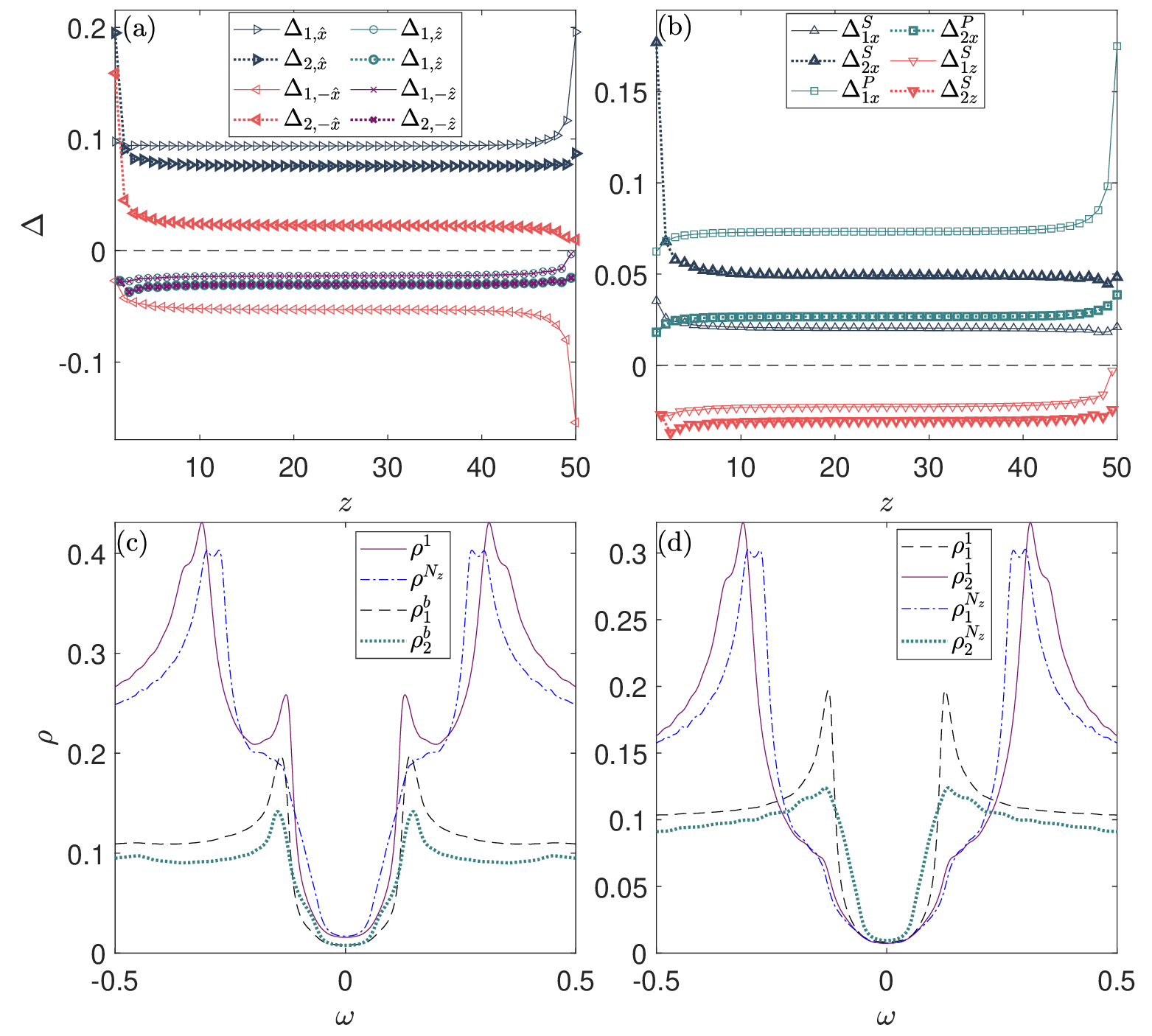}
\caption{\label{Fig2} Superconducting pairing and superconducting-state electronic structure with partially opening boundary condition in the $z$ direction. Top row: the layer-dependent superconducting order parameters obtained by the self-consistent method. Bottom row: orbital-resolved superconducting-state LDOS curves on the $z = 1$, $z = N_z$ surfaces, and in the bulk ($z = N_z/2$).}
\end{figure}

\begin{figure*}
\centering
\includegraphics[width = 18cm]{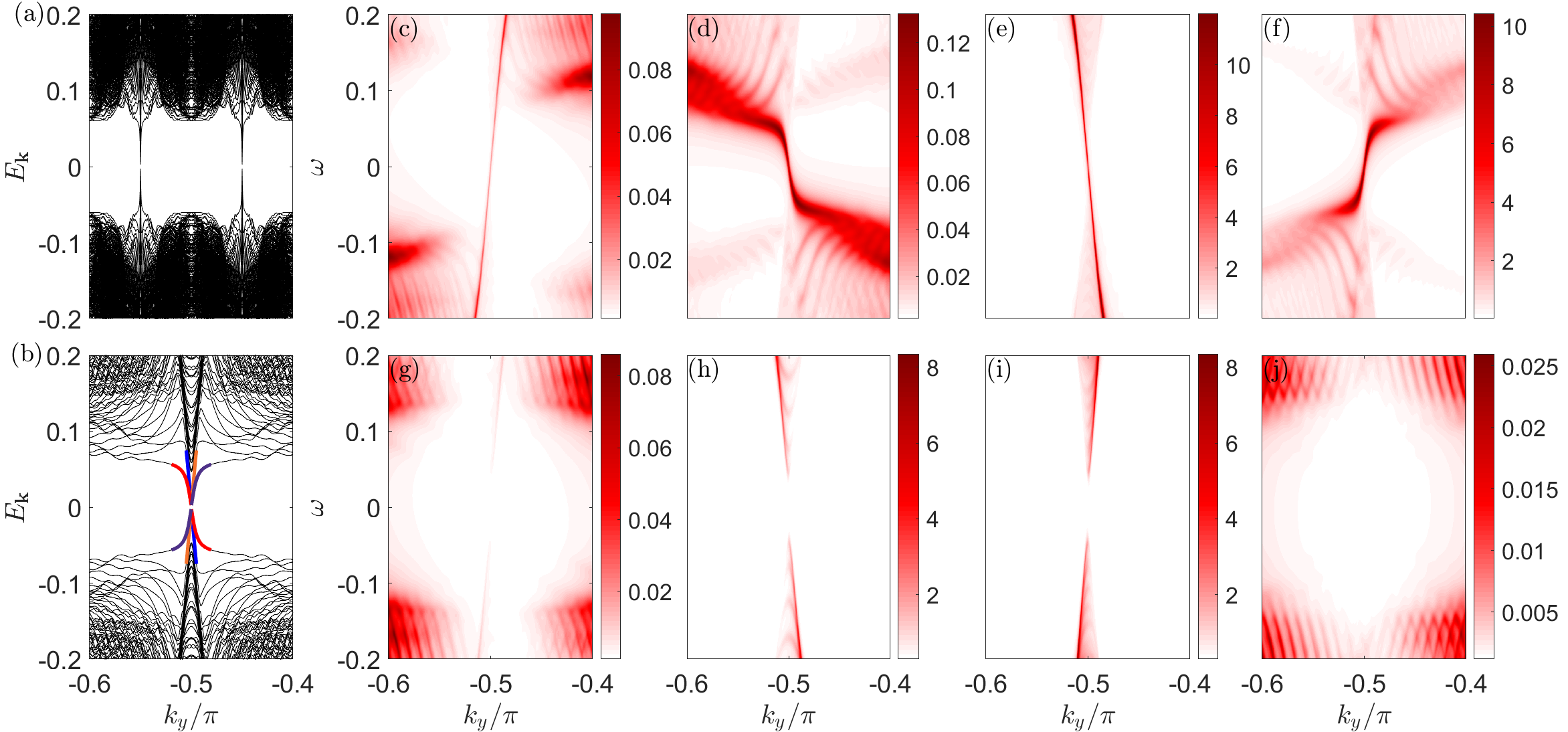}
\caption{\label{Fig3} Electronic structures with partially opening boundary condition in the $x$ and $z$ directions.
In the left panel, (a) superconducting spectrum; (b) zoom-in of the region of the left edge bands.
In the right panel, orbital dependent spectral functions at the four corners (c–f) and at the lateral edges (g–j).}
\end{figure*}

\subsection{Hybrid pairing in the superconducting state}
Fig.~\ref{Fig2} presents self-consistently determined superconducting pairing and the resulting LDOS spectra in the superconducting state with partially opening boundary condition in the $z$ direction. Utilizing the self-consistent method in EQs.[4,5], the superconducting pairing order parameters at each layer can be obtained.
In Fig.~\ref{Fig2}(a), the order parameters are large at the edge, then quickly decrease to constants from the edge into the bulk, owing to numerous zero-energy electronic states contributed by Fermi arcs localized in the surfaces. Both $\Delta_{1,\hat{z}} = \Delta_{1,-\hat{z}}$ and $\Delta_{2,\hat{z}} = \Delta_{2,-\hat{z}}$ in each layer indicate only singlet pairings in the $z$ direction, while $\Delta_{1,\hat{x}} \neq \pm\Delta_{1,-\hat{x}}$ and $\Delta_{2,\hat{x}} \neq \pm\Delta_{2,-\hat{x}}$ imply that both singlet and triplet pairings emerge in the $x$ direction. In the $y$ direction, the order parameters are small and neglected here.

To be clear, these order parameters can be modified into the familiar $s$-wave, $p$-wave forms. We obtain the extended $s$-wave order parameters as $\Delta_{\tau x}^{S}\left(z\right) = \left( \Delta_{\tau,\hat{x}}^z + \Delta_{\tau,-\hat{x}}^z \right)/2$, and $\Delta_{\tau z}^{S}\left(z,z'\right) = \left( \Delta_{\tau}^{z,z'} + \Delta_{\tau}^{z',z} \right)/2$. The triplet $p_x$-wave order parameters can be obtained by $\Delta_{\tau x}^{P}\left(z\right) = \left( \Delta_{\tau,\hat{x}}^z - \Delta_{\tau,-\hat{x}}^z \right)/2$. Then the $s$-wave and $p$-wave form order parameters are depicted in Fig.~\ref{Fig2}(b).
In Fig.~\ref{Fig2}(b), the intralayer $s$-wave order parameter for orbital $2$ $\left(\Delta_{2x}^{S}\right)$ is dominant on the $z = 1$ surface, while $p_x$-wave order parameter for orbital $1$ $\left(\Delta_{1x}^{P}\right)$ is dominant on the $z = N_z$ surface, due to the different normal-state electron structures on these two surfaces mentioned in Fig.~\ref{Fig1}.
Besides, the intralayer $s$-wave order parameter for orbital $1$ $\left(\Delta_{1x}^{S}\right)$, $p_x$-wave order parameter for orbital $2$ $\left(\Delta_{2x}^{P}\right)$ and interlayer $s$-wave order parameters $\left(\Delta_{\tau z}^{S}\right)$ for both two orbitals are weaker and almost maintain a small range both in bulk and surfaces.

The LDOS curves of the superconducting states are shown in Figs.\ref{Fig2}(c) and \ref{Fig2}(d). All the LDOS curves near zero energy are $\cup$-shaped, revealing the system is fully gapped in the superconducting states. In Fig.~\ref{Fig2}(c), the bulk LDOS has a single smaller superconducting gap, while the surface LDOS possesses a two-gap feature. For surface LDOS, the higher superconducting coherent peaks, corresponding to the larger gap, are induced by the superconducting Fermi arcs. The location of the smaller superconducting gap nearly overlaps with that in the bulk LDOS, because the smaller gap is the projection of that in the bulk.
In Fig.~\ref{Fig2}(d), for the $z = 1$ surface, the larger gap is dominated by orbital 2 and the smaller gap is the projection of the bulk mainly for orbital 1. Conversely, on the $z = N_z$ surface, the larger and smaller gaps are dominated by orbital 1 and 2, respectively.

Spatial distinction of the pairing order on the opposite surfaces is rooted in the different Fermi-arc geometries. In the ${k_x}$-$k_y$ plane, extended $s$-wave pairing varies as $\cos {k_x}$ or $\cos {k_y}$ vanishes at ${k_x} = \pm\pi/2$ or ${k_y} = \pm\pi/2$, whereas triplet $p$-wave pairing varies as $\sin {k_x}$ or $\sin {k_y}$ is maximal there. Both in the two surfaces, the horizontal Fermi arcs are localized at ${k_y} = \pm 0.5\pi$, resulting in the weakness of $s_y$-wave pairing. The $z = 1$ surface hosts vertical Fermi arcs near ${k_x} = 0$ line, which stabilize a dominant $s_x$-wave pairing with a secondary $p_x$-wave component. The opposite situation occurs on the $z = N_z$ surface, where vertical arcs near ${k_y} = \pm\pi/2$ line lead to a predominant $p_x$-wave pairing with a subsidiary $s_x$-wave component. The hybrid pairing of $s_x$- and $p_x$-wave can make the maximal gap at the position of the vertical Fermi arcs and fully gap the system.

Moreover, the $p_y$ component of triplet pairing is absent. On a square lattice $p_x$ and $p_y$ pairing usually coexist and can combine into a time-reversal-breaking $p_x + ip_y$ ($p+ip$-wave) state that strongly competes with singlet pairing. In our model, however, the $C_4$ symmetry is absent and the Fermi surface passes through $(0, 0)$ and $(\pm\pi, \pm\pi)$—the nodal points of a $p + ip$ gap—so a pure $p + ip$ state would leave the system gapless and cannot be the thermodynamic ground state; self-consistent calculations confirm that this channel is never favored. As illustrated in Fig. \ref{Fig1}, the normal-state Fermi arcs are predominantly $k_x =$ constant lines that traverse the entire Brillouin zone, whereas $k_y$ = constant arcs are short and fragmentary; this anisotropy stabilizes $p_x$ and $s_x$ pairing over $p_y$ and $s_y$ pairing.

Thus, in TRWSM systems, the geometry of Fermi surface, formed by Fermi arcs localized on the surfaces, is one of significant factors in determining the preferred superconducting pairing symmetry. Searching for or engineering TRWSM with specific Fermi-arcs configurations presents a viable pathway to achieve unconventional superconductivity in these materials.

\subsection{Topological properties}
To interrogate the second-order topology of the superconducting phase we impose partially opening boundaries condition in both $x$ and $z$ directions. The electronic structures with this boundary condition are shown in Fig.~\ref{Fig3}.
The energy band is presented in Fig.~\ref{Fig3}(a), and is symmetric about ${k_y} = 0$. For clarity, the enlarged view for the left edge bands is shown in Fig.~\ref{Fig3}(b). In Figs.~\ref{Fig3}(a) and \ref{Fig3}(b), the bulk energy band is fully gapped, and the edge bands are represented by the coloured lines, which intersect at ${k_y} = \pm\pi/2$. This indicates the non-trivial second-order topology in the system.

The spectral functions at four corner sites are exhibited in Figs.~\ref{Fig3}(c)-\ref{Fig3}(f). In these figures, the ascending edge band with the orange line can be identified by spectral function $A_1^{\left( 1,1 \right)}$, as shown in Fig.~\ref{Fig3}(c). The descending edge band with the red line can be recognized by spectral function $A_2^{\left( 1,N_z \right)}$, being presented in Fig.~\ref{Fig3}(d). The spectral functions for both orbitals are notably weak since orbital 1 and orbital 2 have minimal weight on the $z = 1$ and $z = N_z$ surfaces, respectively.

The descending edge band with the blue line and the ascending edge band with the purple line can be well identified by spectral function $A_2^{\left( N_x,1 \right)}$ [Fig.~\ref{Fig3}(e)] and $A_1^{\left( N_x,N_z \right)}$ [Fig.~\ref{Fig3}(f)], respectively. The last two spectral functions are much larger than the first two. These spectral functions for the four corner edge bands reveal that the mirror symmetry in $x$-$z$ plane has been broken in the superconducting states. The proportion of the four edge states is in extreme disequilibrium.

Figs.~\ref{Fig3}(g)-\ref{Fig3}(j) depict the spectral functions at the middle of the $z = 1$ and $z = N_z$ surfaces. In Figs.~\ref{Fig3}(g) and ~\ref{Fig3}(h), the spectral functions $A_1^{\left( N_x/2,1 \right)}$ and $A_2^{\left( N_x/2,1 \right)}$ are shown, while the spectral functions $A_1^{\left( N_x/2,N_z \right)}$ and $A_2^{\left( N_x/2,N_z \right)}$ are presented in Figs.~\ref{Fig3}(i) and ~\ref{Fig3}(j). All these four spectral functions reveal that the Fermi arcs have been fully gapped by the hybrid superconducting pairing and the topological edge states are only localized at the four hinges.

\begin{figure}
\centering
\includegraphics[width = 6cm]{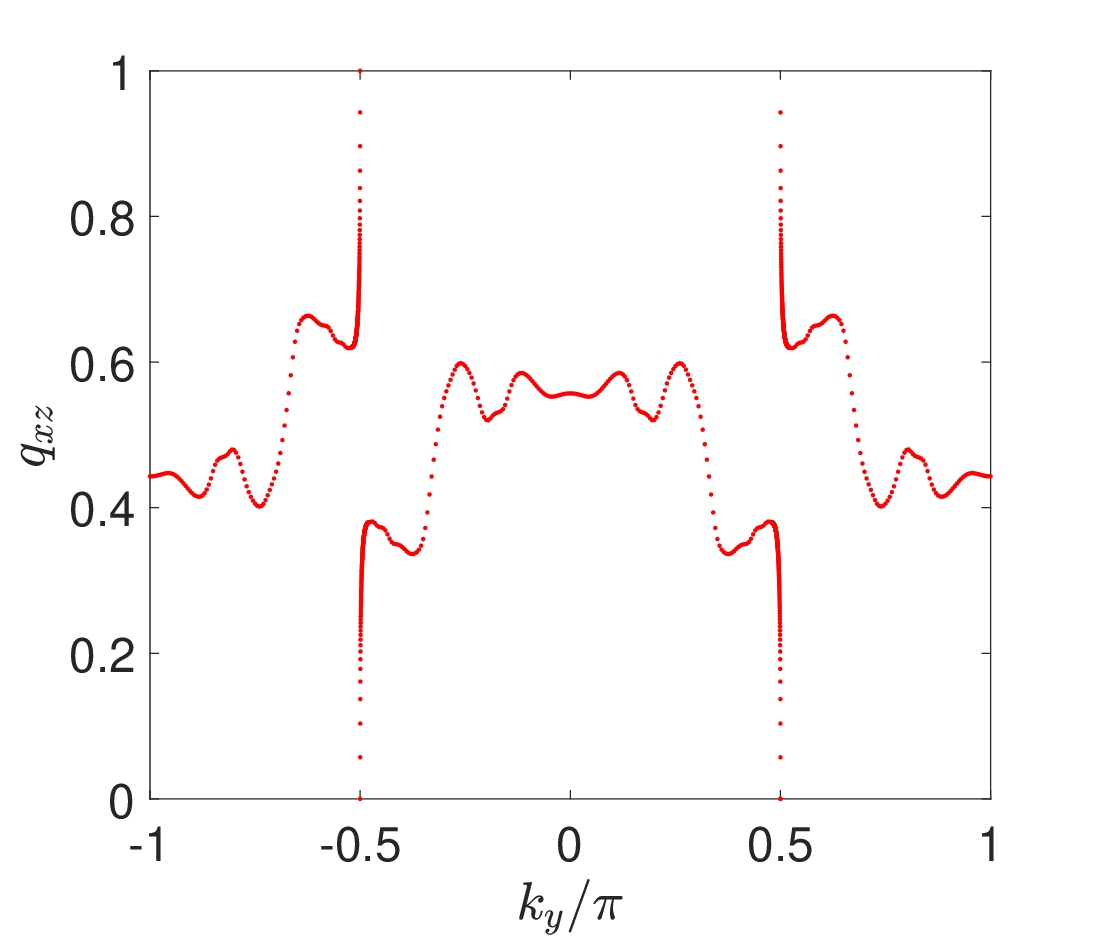}
\caption{\label{Fig4} Quadrupole moment for each ${k_y}$-sliced layer.}
\end{figure}

To confirm the higher-order topology further we compute the quadrupole moment $q_{xz}\left( {k_y} \right)$ [Eq.(8)]. The quadrupole moment $q_{xz}\left( {k_y} \right)$ for each ${k_y}$-sliced layer can be utilized to describe the non-trivial second-order topology in the system. The curve of quadrupole moment is shown in Fig.~\ref{Fig4}. Time-reversal symmetry enforces $q_{xz}\left(k_y\right) = q_{xz}\left(-k_y\right)$. In Fig.~\ref{Fig4}, the quadrupole moment has quite sudden changes and makes windings at ${k_y} = \pm\pi/2$, indicating the non-trivial second-order topology of the system. The windings of quadrupole moment at ${k_y} = \pm\pi/2$ are relevant to the edge-band crossings at ${k_y} = \pm\pi/2$. Besides, the windings of quadrupole moment at ${k_y} = -\pi/2$ and ${k_y} = \pi/2$ are in the opposite direction, implying that the hinge states here are not chiral and without any current passing through in our system and TR breaking.

Our study positions type-II TRWSM as a programmable laboratory for on-demand unconventional superconductivity. Multiple momentum-resolved Fermi arcs coexist on each surface, their distinct orbital textures and connectivities acting as built-in selectors that stabilize singlet or triplet pairing on opposite surfaces without external interfaces or proximity effects. Additionally, type-II TRWSM dispersion carves out an extended Fermi surface that hosts a large density of states at the Fermi level; the profusion of zero-energy carriers cooperatively amplifies the pairing instability and points toward a realistic route to push the critical temperature of topological superconductivity well beyond the ceilings of type-I or TR-breaking counterparts.

Notably, the positions of Weyl points and their associated Fermi arcs can be reconfigured through surface decoration \cite{yang_topological_2019}, pressure \cite{cheng_tunable_2024}, or strain \cite{sie_ultrafast_2019}. This tunability positions type-II TRWSMs as a versatile single-material arena, where high‑$T_c$ unconventional superconductivity, nontrivial topology, and other emergent quantum states can be generated, controlled, and hybridized. Such a system provides an exceptional setting for exploring novel and hybrid quantum phases.

\section{summary}
In summary, we have employed a minimal model for type-II TRWSM. In this system, a large Fermi surface is formed by bulk electron and hole pockets, complemented by the Fermi arcs localized on the surfaces. The Fermi pockets supply mainly bulk zero-energy electronic states, while the Fermi arcs induce a pronounced zero-energy peak in the LDOS on the surfaces. Through the self-consistent calculation, a mixed $s$-wave and $p$-wave pairing is revealed and coexists both in bulk and on the surfaces. A spatial dichotomy has been examined: $s$-wave pairing is dominant on one surface, whereas the $p$-wave pairing is prominent on the other surface. This distinct pairing mechanism stems from the differing patterns of Fermi arcs on the two surfaces. In WSM, various Fermi-arc configurations may induce an intrinsic unconventional topological superconductivity in the system.

In the superconducting states of our model, mirror-symmetry-breaking hinge states emerge and are dominated by different orbital channels. We have provided numerical evidences for energy bands, spectral functions in the four hinges. Furthermore, the second-order topological property is also corroborated by the windings of quadrupole moment at certain momenta, which is directly linked to the edge-band crossing at these specific momenta. Our findings suggest that type-II TRWSM be a promising platform for exploring unconventional and topological superconductivity.

\begin{acknowledgments}
This work was supported by Guangdong Provincial Quantum Science Strategic Initiative (Grant No. GDZX2404001), the NSFC/RGC JRS grant (N\_HKU 774/21), and GRF of Hong Kong (17303/23).
\end{acknowledgments}

%\bibliography{Weyl.bib}

%apsrev4-2.bst 2019-01-14 (MD) hand-edited version of apsrev4-1.bst
%Control: key (0)
%Control: author (8) initials jnrlst
%Control: editor formatted (1) identically to author
%Control: production of article title (0) allowed
%Control: page (0) single
%Control: year (1) truncated
%Control: production of eprint (0) enabled
%

\end{document}